\documentclass[pra,twocolumn,notitlepage,superscriptaddress,longbibliography]{revtex4-1}
\usepackage{amsmath}
\usepackage{amssymb}
\usepackage{mathrsfs}
\usepackage[english]{babel}

\usepackage[unicode=true,colorlinks=true,citecolor=blue,urlcolor=blue]{hyperref}

\usepackage{bm}
\usepackage{color}
\usepackage{epsfig}

\usepackage[normalem]{ulem}

\renewcommand {\i}{{\rm i}}
\renewcommand {\phi}{{\varphi}}

\begin{document}

\title{Polaron effect in waveguide quantum optomechanics}

\author{Denis Ilin}
\affiliation{School of Mathematical and Physical Sciences, University of Technology Sydney, Ultimo, NSW 2007, Australia}\affiliation{ Sydney Quantum Academy, Sydney, NSW 2000, Australia}

\author{Alexander S. Solntsev}
\affiliation{School of Mathematical and Physical Sciences, University of Technology Sydney, Ultimo, NSW 2007, Australia}

\author{Ivan Iorsh}
\affiliation{Queen's University, Kingston, Canada}

\begin{abstract}
We investigate the impact of the quantized mechanical motion of optically trapped atoms, arranged in proximity to a one-dimensional waveguide, on the propagation of polariton modes. Our study identifies a regime of resonant phonon-assisted mixing between lower and upper polaritons, resulting in a pronounced polaron effect. This effect is characterized by the formation of new band gaps and the appearance of weakly dispersive states within the original polariton band gap. The polaron spectrum, which can be directly probed via resonant elastic scattering, provides novel opportunities for quantum optical applications. These findings open avenues for enhanced control in state-of-the-art waveguide quantum electrodynamics experiments with cold atoms.

\end{abstract}

\maketitle

\section{Introduction}
The area of waveguide quantum electrodynamics (WQED)~\cite{Roy2017,KimbleRMP2018,RevModPhys.95.015002} studies the propagation of light through one-dimensional arrays of quantum emitters. The exchange of photons between quantum emitters facilitates the emergence of the long-range quantum correlations between quantum emitters, which in turn lead to the multiphoton quantum correlations in the light scattered off these structures. Over the past decade, advances in technology have enabled the realization of WQED systems on various platforms, such as cold atoms~\cite{PhysRevLett.117.133603,Corzo2019,menon2024integrated,PhysRevLett.128.073601,hattermann2017coupling}, semiconductor quantum dots~\cite{foster2019tunable,jalali2020chiral,PhysRevLett.131.033606,uppu2020chip,mccaw2024reconfigurable} and superconducting qubits~\cite{vanLoo2013,Mirhosseini2019,kannan2020waveguide,kannan2023demand,kockum2018decoherence}. 
These developments have catalyzed significant research, exploring fundamental quantum phenomena such as quantum chaos~\cite{PhysRevLett.126.203602}, quantum spin glass phases~\cite{kroeze2023replicasymmetrybreakingquantumoptical}, and practical applications in quantum memory~\cite{PhysRevLett.114.180503} and quantum information processing~\cite{guimond2020unidirectional,PhysRevLett.130.023601}.

One of the emerging directions in the studies of the WQED systems is the optomechanical interactions in these systems. The mechanical motion of the optically trapped cold atoms is quantized at sufficiently low temperature, and the scattering of the photon off the atom may induce transition between the mechanical motion eigenstates. The efficiency of these processes is defined by the Lamb-Dicke parameter which is the ratio between the kinetic energy transferred to the atom by the scattered photon and the phonon energy in the optical trap~\cite{castin1991quantization}. This parameter can be flexibly tuned by varying the optical trap profile, and thus the optomechanical interaction can be altered in a wide range. The optomechanical interaction in WQED systems was shown to lead to a plethora of fascinating effects ranging from the self-organisation of atoms~\cite{chang2013self} to the emergence of the highly entangled phases~\cite{manzoni2017designing, iorsh2020waveguide, sedov2020chiral, PhysRevB.109.165419, arguello2022optomechanical, weng2022exact}, where the atomic degrees of freedom and the mechanical motion of the atoms are strongly correlated.

Most of the studies so far have focused on the systems comprising a small number of qubits, mainly due to the fact that the description of the correlated dynamics of photons, phonons and qubit excitations is computationally challenging, mainly due to the highly nonlinear form of the effective atom-phonon interaction.  At the same time, in recent experiments, the arrays of more than 100 atoms have been demonstrated, where the formation and propagation of the hybrid light-matter wave, polariton could be directly probed~\cite{Corzo2019}. Thus, it is natural to investigate the role of the optomechanical interaction in the dispersion and decay properties of the polariton state. Specifically, it is tempting to understand what the properties of the polaron states are, i.e. polaritons dressed in phonon clouds in these systems. Moreover, it is not clear if the atom-phonon interactions can ultimately lead to the polaron localization, which could be interesting both as a fundamental physical effect and for the possible applications in quantum optical information storage. 
\begin{figure}[t]
\centering
\includegraphics[width=0.45\textwidth]{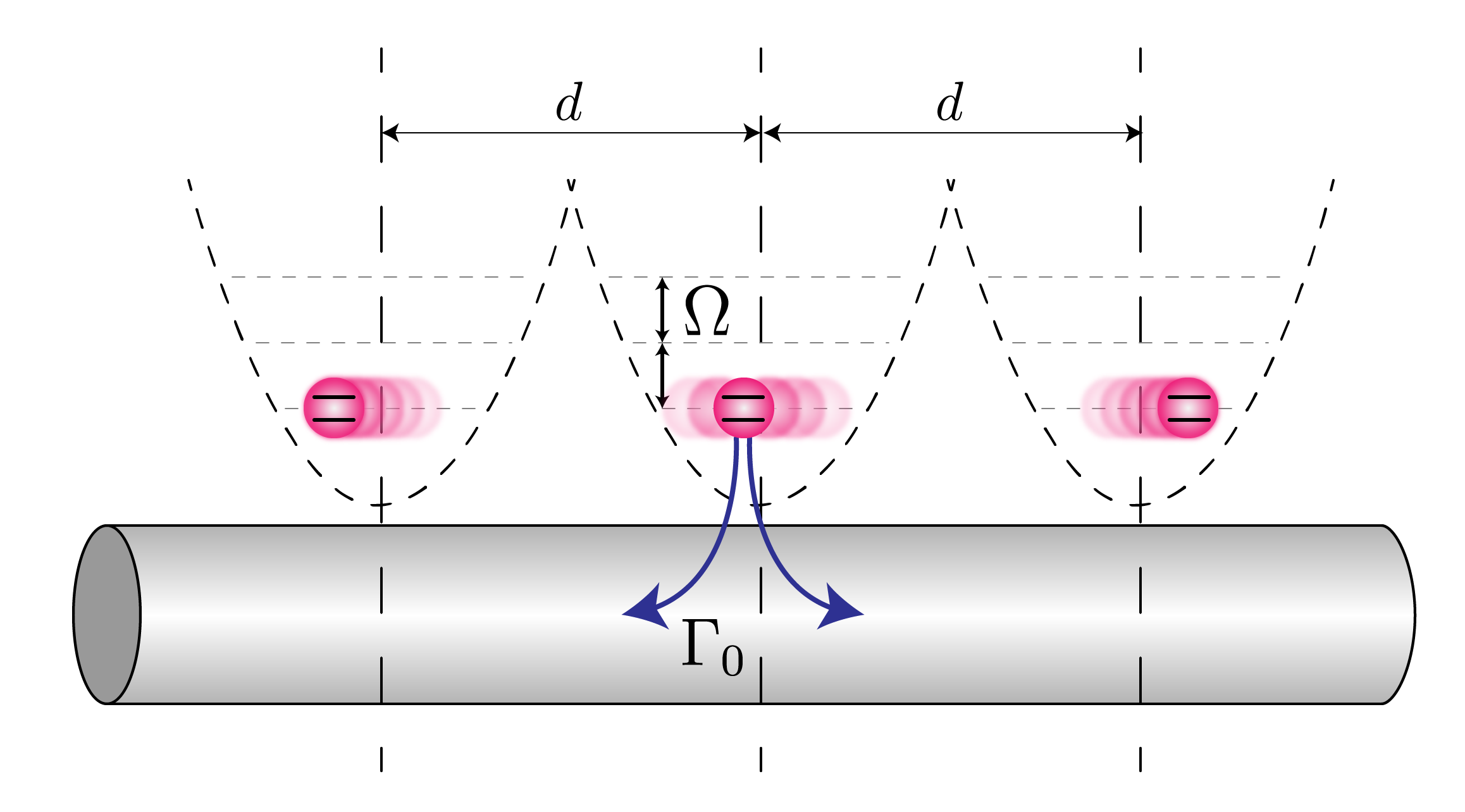}
\caption{Schematic image of the considered system. An array of two level systems in optical traps interact with each other via the exchange of the waveguide photons.}\label{fig:10}
\end{figure}

In this Article, we develop a formalism which allows one to find the polaron dispersion of the WQED setup with optomechanical interaction of arbitrary strength shown in Fig.~\ref{fig:10}. The formalism is based on the unitary transformation allowing us to map the considered system to analogue of Holstein model with linear exciton-phonon coupling for arbitrary optomechanical strengths. This model can then be analyzed with standard numerical techniques such as Momentum Average Approximation (MAA)~\cite{berciu2007systematic}. We show that the optomechanical interaction leads to the formation of slow-light bands inside the polaritonic band gap of the atomic array as well as to the formation of additional band gaps corresponding to the phonon-mediated anticrossing of the lower and upper polaritonic bands.

\section{Model}
We start from the Hamiltonian describing the WQED system with optomechanical interaction: an array of identical two-level systems (qubits)  trapped by the parabolic potentials which interact with photonic modes in waveguide. The  Hamiltonian of the problem is 
\begin{equation}
    \begin{aligned}
        \hat{H} &= \omega_0\sum\limits_{j=1}^{N}\hat{\sigma}^{\dagger}_j\hat{\sigma}_j + \sum\limits_{k}\omega_k\hat{a}^{\dagger}_k\hat{a}_k\\&+\sum\limits_{j=1}^{N}\frac{\hat{p}_{j}^2}{2M}+\frac{M\Omega^2 \hat{z}_j^2}{2}\\&-\frac{g}{\sqrt{L}}\sum\limits_{j=1}^{N}\sum\limits_{k}\left[\hat{\sigma}^{\dagger}_j\hat{a}_ke^{ik(dj+\hat{z}_j)}+h.c.\right] \label{eq:H0}
    \end{aligned}
\end{equation}
where $\omega_0$ is the resonance frequency of qubits, $g$ is the strength of qubit-photon interaction, $\hat{\sigma}^{\dagger}_{j}(\hat{\sigma}_{j})$ are rising (lowering) operators of $j-$th qubit, $d$ is the period of the structure. Operators $\hat{p}_j,\hat{z}_j$ describe the quantized one-dimensional motion of atoms inside the traps, and $M,\Omega$ are the atoms' mass and the trap resonant frequency. We assume one-dimensional motion along the waveguide, as the atoms are tightly confined in the perpendicular dimensions. The dispersion of photons in waveguide is linear, $\omega_k=c|k|$, and $\hat{a}^{\dagger}_{k}(\hat{a}_{k})$ is creation (annihilation) operator of the photon with momentum $k$. Physically, the electron-phonon interaction term describes the transfer of the momentum $k$ between atom and photon as the photon is either emitted or absorbed. We can introduce creation and annihilation operators for phonons in the $j-$th trap, $\hat{b}_j,\hat{b}^{\dagger}_j$ and $\hat{z}_j=\alpha (\hat{b}_j+\hat{b}^{\dagger}_j)$, where $\alpha=\sqrt{\hbar/(2M\Omega)}$ is characteristic length scale of the atomic motion.

The parameter $\alpha$ remains small in typical experimental setups. Indeed, if we take the lithium atoms then  $\omega_0\sim5\times10^{14}$Hz. The characteristic length scale of the optical trap $\lambda$ is approximately $100$ nm. Hence, the corresponding phonon energy is about $\Omega\sim 3.5\times10^{5}$~Hz and consequently $\alpha_0=\omega_0\alpha/c\approx 0.1$, where $c$ is speed of light in the waveguide. This corresponds to a regime where the characteristic uncertainty in the atom’s position within the trap is much smaller than the wavelength of light at the resonant frequency. Moreover, the light-matter interaction is enhanced only in the vicinity of the resonant frequency. We are considering the weak light-matter coupling, and thus we approximate $k\alpha\approx k_0 \alpha$, where $k_0=\omega_0/c$. The reason for that is the fact that the residual term $(k-k_0)\alpha$ contains both small fluctuations of momentum around resonance and small fluctuations of atoms' positions, and thus can be neglected. Within this approximation we can apply a unitary transformation: 
\begin{align}
\hat{U}=\exp\left[ik_0\sum\limits_{j=1}^{N}\hat{\sigma}^{\dagger}_j\hat{\sigma}_j\hat{z}_j\right]. \label{eq:U}
\end{align}
This transformation physically moves the system to the moving frames of atoms and is similar to the Lee-Low-Pines transformation. The transformed Hamiltonian reads
\begin{equation}
\begin{aligned}
    \hat{U}^{\dagger}\hat{H}\hat{U} &=\hat{H}_{TLS}+\hat{H}_{ph}-\frac{g}{\sqrt{L}}\sum\limits_{k,j}\left[\hat{\sigma}^{\dagger}_j\hat{a}_ke^{ikdj}+h.c.\right] \\ & +\sum\limits_{j=1}^{N}\frac{\left(\hat{p}_{j}+k_0\hat{\sigma}^{\dagger}_j\hat{\sigma}_j\right)^2}{2M}+\frac{M\Omega^2 \hat{z}_j^2}{2}, \label{eq:H_trans1}
\end{aligned}
\end{equation}
where $\hat{H}_{TLS}=\omega_0\sum\limits_{j=1}^{N}\hat{\sigma}^{\dagger}_j\hat{\sigma}_j$, and $H_{ph}=\sum\limits_{k}\omega_k\hat{a}^{\dagger}_k\hat{a}_k$. We note that after the transformation only qubit degrees of freedom are coupled to phonons explicitly.

The first line of the expression represents nothing but the Dicke model in the presence of weak light-matter coupling. We can therefore integrate out the photon degrees of freedom~\cite{RevModPhys.95.015002} and obtain the effective Hamiltonian
\begin{equation}
\begin{aligned}
\hat{H}&=\omega_0\sum_{j}\hat{\sigma}_j^{\dagger}\hat{\sigma}_j-i\Gamma_0\sum_{j,l}\hat{\sigma}_{j}^{\dagger}\hat{\sigma}_l e^{i\varphi |j-l|} \\&+\hat{H}_{TLS-phon}, \label{eq:H_trans2}
\end{aligned}
\end{equation}
where $\Gamma_0=g^2/c$ is the radiative lifetime of the qubit, $\varphi=k_0d$, and $\hat{H}_{TLS-phon}$ is the second line of Eq.~\eqref{eq:H_trans1}. Hamiltonian~\eqref{eq:H_trans2} is non-hermitian due to the leakage of excitations via the ends of the finite array. At the same time, total number of excitations $\sum_{j}\hat{\sigma}_j^{\dagger}\hat{\sigma}_j$ is a good quantum number. In what follows, we consider the case of only a single excitation. We note that for only two qubits the Hamiltonian~\eqref{eq:H_trans2} is equivalent to the optomechanical Hamiltonian derived in~\cite{iorsh2020waveguide} by other methods

To consider the case of infinite arrays, we make the Fourier transform and subsequent rotation $\hat{b}_k\rightarrow i\hat{b}_k$. Finally, for the case of just a single qubit excitations the spin operators can be changed to the bosonic creation and annihilation operators $\hat{\sigma}_q\rightarrow \hat{c}_q$, and the effective Hamiltonian yields
\begin{equation}
    \begin{aligned}        \hat{H}_{\text{eff}}=&\sum\limits_{q}\left[\epsilon(q)\hat{c}^{\dagger}_q\hat{c}_q+\Omega\hat{b}^{\dagger}_q\hat{b}_q+\frac{\beta}{\sqrt{N}}\sum\limits_{k}\hat{c}^{\dagger}_{q-k}\hat{c}_q(\hat{b}^{\dagger}_k+\hat{b}_{-k})\right], \label{eq:H_fin}
    \end{aligned}
\end{equation}
where
\begin{align}
\epsilon(q)=\omega_0+\Gamma_0\frac{\sin(\varphi)}{\cos(qd)-\cos(\varphi)} \label{eq:disp}
\end{align}
is the dispersion law for the infinite qubit array,  and $\beta=\Omega\alpha_0$ is the effective phonon-polariton coupling strength.  The Hamiltonian~\eqref{eq:H_fin} is a direct analogue of the Holstein polaron problem with the modified particle dispersion relation $\epsilon(q)$. It is known that there are no self-trapping (localization) phase transitions in the electron-phonon Hamiltonians of this type~\cite{lowen1988absence} unless $\Omega>0$. In the case of $\Omega\rightarrow0$ and $M\Omega^2\rightarrow \kappa >0$ the system described by Hamiltonian~\eqref{eq:H0} can be treated classically as an array of resonant scatterers with random distances. Indeed, in this limit atoms are not moving while emitting or absorbing photons,  and thus the positions of atoms are fixed. The positions of atoms are distributed according to Boltzmann distribution with probabilities $p_i\sim e^{-\kappa z^2_j/kT}=e^{-z^2_j/\sigma^2}$, where $T$ is the chain temperature. As we show in the \textbf{Appendix} using transfer matrix approach, in this case the localization is always present even for arbitrarily small fluctuations.

In the case of finite phonon frequencies, translational invariance is preserved and the momentum $q$ is a good quantum number. To find the eigenstates of the Hamiltonian~\eqref{eq:H_fin} we use the Momentum Average Approximation (MAA)~\cite{goodvin2006green}. Within MAA, we look for the solution for the Green's function $G(\omega, k)=\langle 0|\hat{c}_k [\omega-\hat{H}_{eff}]^{-1}\hat{c}_{k}^{\dagger}|0\rangle$, where $\hat{H}_{eff}$ is given by eq.~\eqref{eq:H_fin} and $|0\rangle$ is the polariton and phonon vacuum state.   Dyson series for the Green's function is expressed as an infinite series of coupled equations for the auxiliary functions $F_n$:
\begin{align}
&F_n(k,q_1\ldots q_n,\omega)=\nonumber \\& \langle 0|\hat{c}_k [\omega-\hat{H}_{eff}]^{-1}\hat{c}_{k-\sum_{i=1}^{n}q_i}^{\dagger}\hat{b}^{\dagger}_{q_1}\ldots \hat{b}^{\dagger}_{q_n} |0\rangle.
\end{align}
The essence of the MAA$^{(n)}$ approximation is to average over the phonon momenta for all $n'>n$, which results in the linear system of equations for $\tilde{F}_n'(k,\omega)={N}^{-n'}\int\Pi_{i=1}^{n'} dq_i F_n'$. This system of equations can be solved exactly and the Green's function $G(k,\omega)=\tilde{F}_0$. MAA allows to take into account all powers of the electron-phonon interaction at the expense of omitting the correlations between the phonons. The polaron Green's function can be written as:
\begin{align}
  G(\omega, q) = \frac{1}{\omega-\epsilon(q)-\Sigma(\omega,q)}, \label{eq:G}
\end{align}
where the self-energy part $\Sigma(\omega,q)$ is momentum independent in the MAA approximation, it happens because the effective phonon-polariton coupling strength is momentum independent, and is expressed via a continuous fraction~\cite{goodvin2006green}:
\begin{align}
\Sigma(\omega)=\dfrac{\beta^2 \bar{g}_0(\omega-\Omega)}{1-\dfrac{2\beta^2\bar{g}_0(\omega-\Omega)\bar{g}_0(\omega-2\Omega)}{1-\dfrac{3\beta^2\bar{g}_0(\omega-2\Omega)\bar{g}_0(\omega-3\Omega)}{1-\ldots}}}, \label{eq:Sigma}
\end{align}
where $\bar{g}_0(\omega)$ is the momentum averaged free polariton  Green's function
\begin{align}
\bar{g}_0(\omega)=\int \frac{dq}{\omega_+-\epsilon(q)}=\frac{1}{\omega_+}-\frac{\omega_+^{-1}}{\sqrt{1-\frac{\omega_+^2}{\Gamma_0^2}+\frac{2\omega_+\cot\varphi}{\Gamma_0}}}, \label{eq:g0}
\end{align}
where $\omega_+=\omega+i\eta$. Equations~\eqref{eq:Sigma},\eqref{eq:G},\eqref{eq:g0} can be used to find the polaron spectra of the system.
\section{Results and Discussion}

\begin{figure}[t]
\centering
\includegraphics[width=0.48\textwidth]{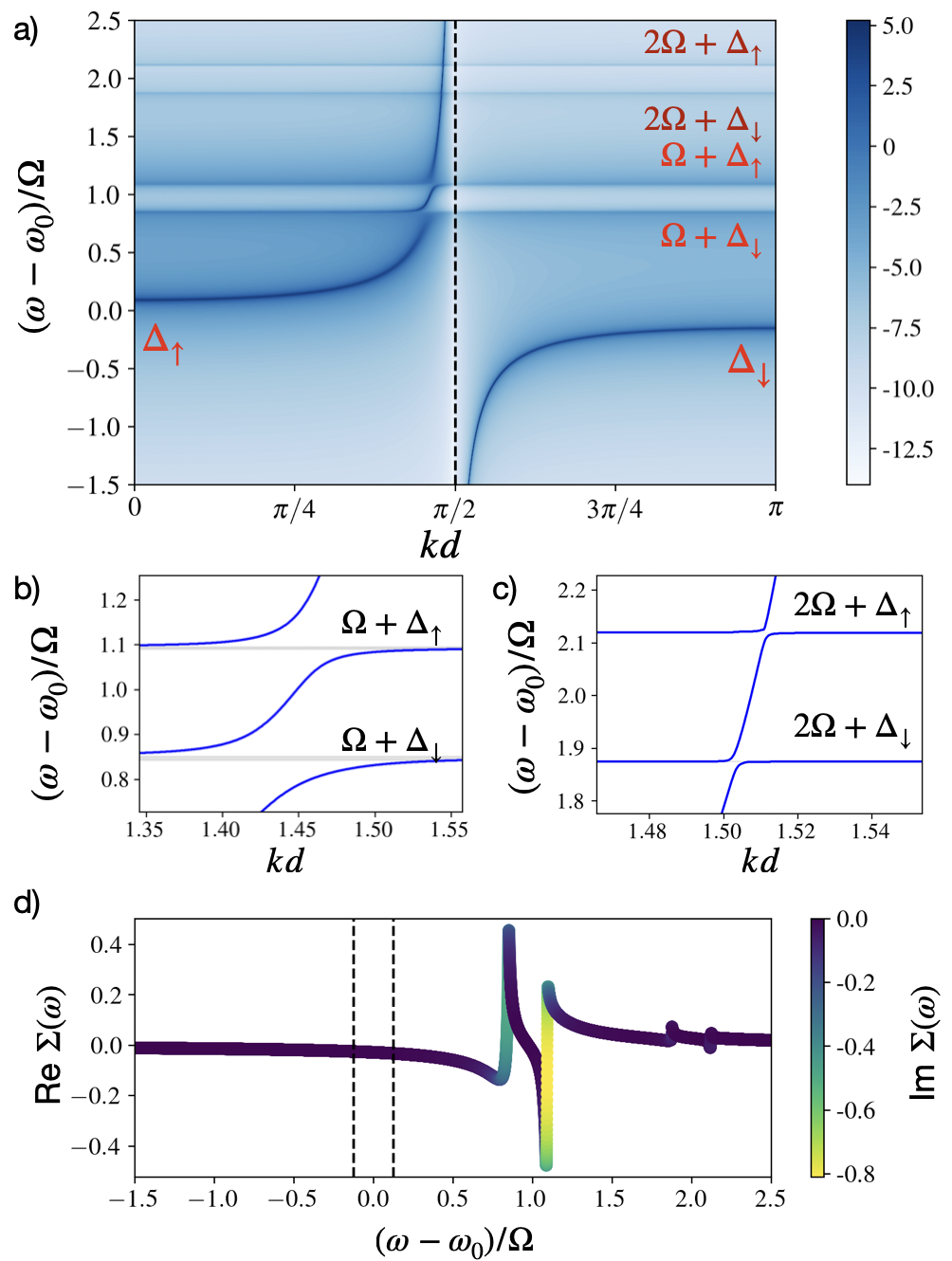}
\caption{The case $\Gamma_0/\Omega=1/8$ and distance $\varphi=\pi/2$ is considered. a) spectral weights $\ln(A(\omega, k))$ in the first Brillouin zone for the model are plotted via $\text{MAA}$. The black dashed line corresponds the divergent point of the polaritonic dispersion. The grey zones denote forbidden zones. b) dispersion of polarons in the vicinity of one phonon intersection. c) dispersion of polarons in the vicinity of two phonons intersection. The grey lines on b) and c) denote new opened band gaps. d) the real and imaginary parts of self-energy part $\Sigma(\omega)$. Black dashed lines correspond to the polaritonic band gap.}\label{fig:2}
\end{figure}

We show the results for the polaron spectral weight $A(\omega,k)=-{\pi}^{-1}\mathrm{Im}(G)$ in Fig.~\ref{fig:2}. In order to understand the origin of the features in the spectral density, let us first recall the properties of the bare polariton dispersion. Due to the long-range photon-mediated coupling between the qubits and within the Markov approximation neglecting the retardation effects, the polariton spectrum $\epsilon(q)$ in Eq.~\eqref{eq:disp} is unbounded and diverges at $\cos(qd)=\cos(\varphi)$. Moreover, the polariton spectrum is characterized by a polariton gap in the vicinity of the qubit resonant frequency. The gap spans from $\Delta_{\downarrow}=\omega_0-\Gamma_0\tan(\varphi/2)$ to $\Delta_{\uparrow}=\omega_0+\Gamma_0\cot\varphi/2$. When the frequency $\omega$ falls within the gap, the averaged Green's function is purely real. At the same time, the self energy $\Sigma$ in Eq.~\eqref{eq:Sigma} contains the averaged Green's function at infinite series of frequencies $\omega-n\Omega$; therefore, even for the frequencies inside the polaritonic gap, there will be non-vanishing imaginary part of the self-energy and thus finite spectral weight. Furthermore, the averaged Green's function has Van-Hove singularities at the band edges $\Delta_{\uparrow,\downarrow}$ (see Fig.~\ref{fig:2}(d)). Therefore, when for some $n$ the expression $\omega-n\Omega$ is close to one of the band edges, this leads to resonances in the self-energy, which in turn facilitates the emergence of the features marked in Fig.~\ref{fig:2}(a). The anticrossings between the polaron branches clearly seen In Fig.~\ref{fig:2}(b,c) occur because electron-phonon interactions couple the bare upper polariton with the composite state of the phonon and one upper or lower polariton. The anticrossing creates new polaron gaps in the spectrum.

\begin{figure}[t]
\centering
\includegraphics[width=0.48\textwidth]{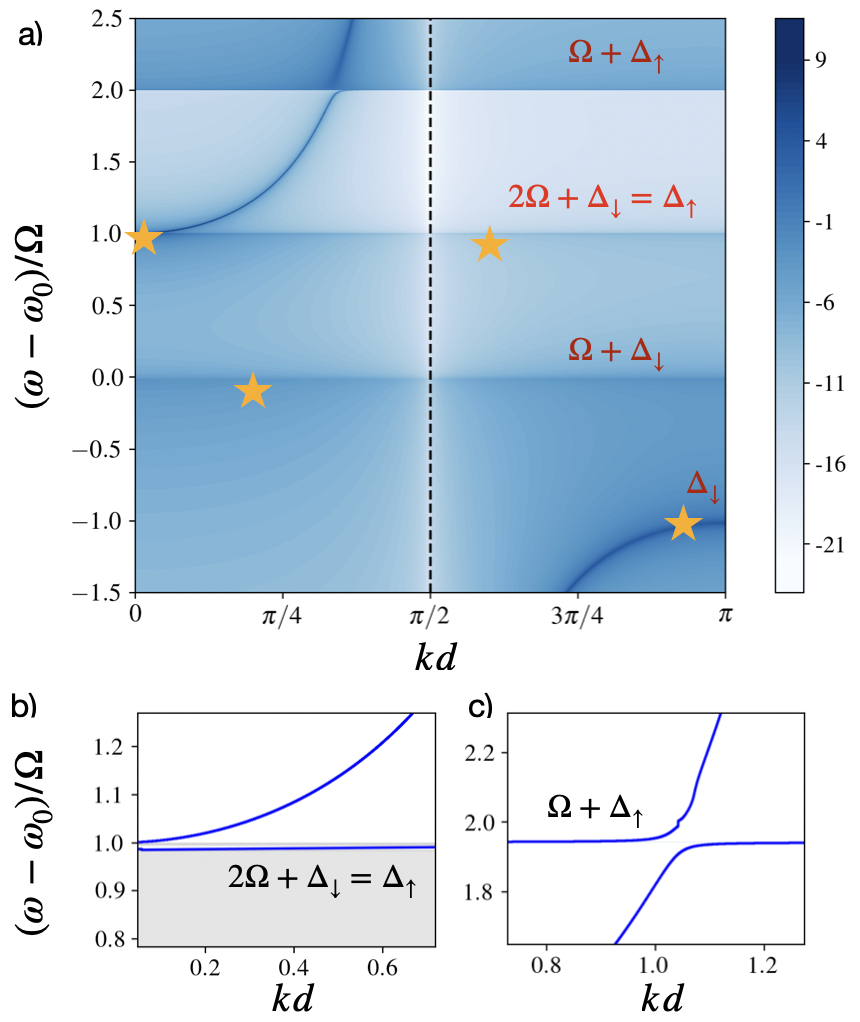}
\caption{The case $\Gamma_0/\Omega=1$ and distance $\varphi=\pi/2$ is considered. a) spectral weights $\ln(A(\omega, k))$ in the first Brillouin zone for the model is plotted via $\text{MAA}$. The black dashed line corresponds to the divergent point of the polaritonic dispersion. The grey zones denote forbidden zones. The yellow stars correspond to the several energy levels in finite $N=6$ system which is modeled numerically. The average number of phonons on these levels matches with its prediction in infinite case. b) dispersion of polarons in the vicinity of one phonon intersection. c) dispersion of polarons in the vicinity of two phonons intersection. The grey lines on b) and c) denote new opened band gaps.}\label{fig:3}
\end{figure}

If we consider the case where the phonon frequency is smaller than the gap width, the bound states of phonon an bare polariton gain spectral weight and lead to the narrow resonances inside the polariton gap, as shown in Fig.~\ref{fig:3}. These resonances should be visible in the reflection spectrum of the system, and thus may be directly optically probed.

\begin{figure}[t]
\centering
\includegraphics[width=0.48\textwidth]{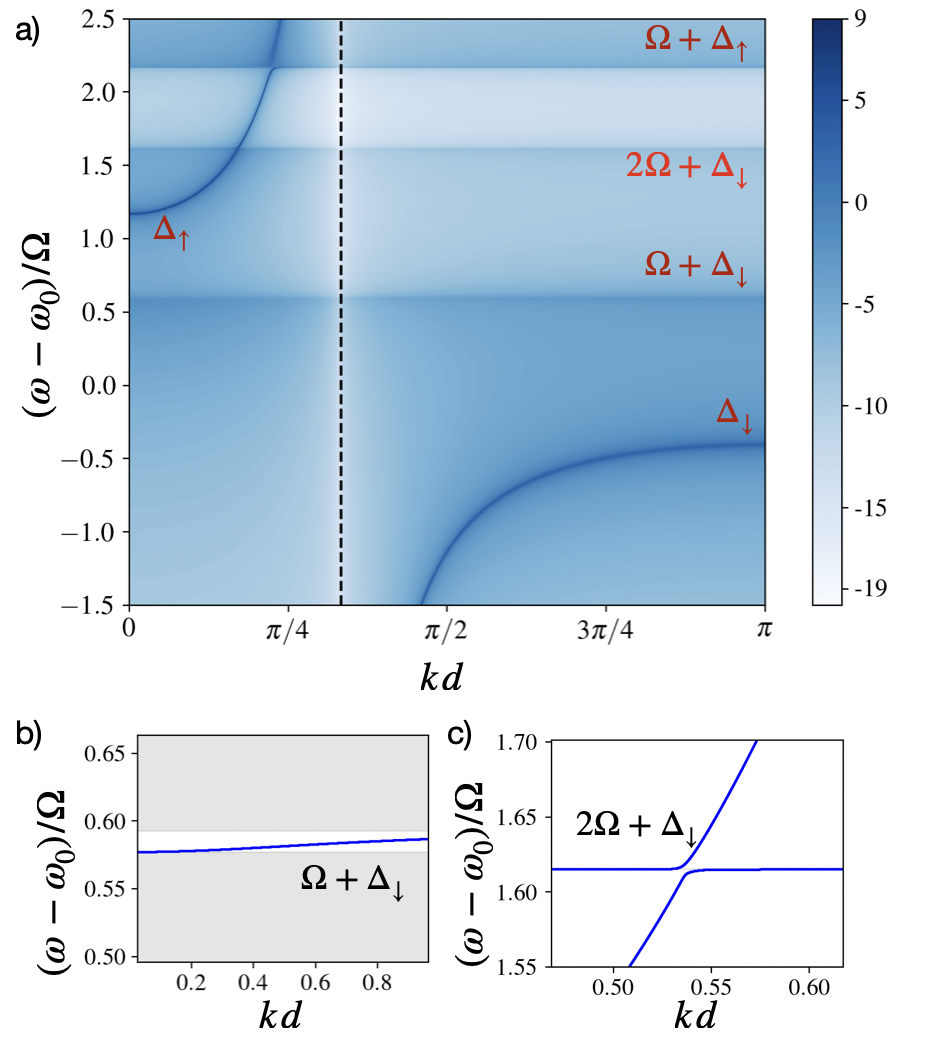}
\caption{The case $\Gamma_0/\Omega=1$ and distance $\varphi=\pi/3$ is considered. a) spectral weights $\ln(A(\omega, k))$ in the first Brillouin zone for the model is plotted via $\text{MA}$. The black dashed line corresponds the divergent point of the polaritonic dispersion. The gray zones denote forbidden zones. b) dispersion of polarons in the vicinity of one phonon intersection. c) dispersion of polarons in the vicinity of two phonons intersection. The grey lines on b) and c) denote new opened band gaps.}\label{fig:4}
\end{figure}

The values of the anticrossing induced gaps and the energies of the in-gap states can be tuned by changing the inter-atomic distance ($\varphi$). In Fig.~\ref{fig:4} we show the spectral weight map for the case of $\varphi=\pi/3$.

\begin{figure}[t]
\centering
\includegraphics[width=0.48\textwidth]{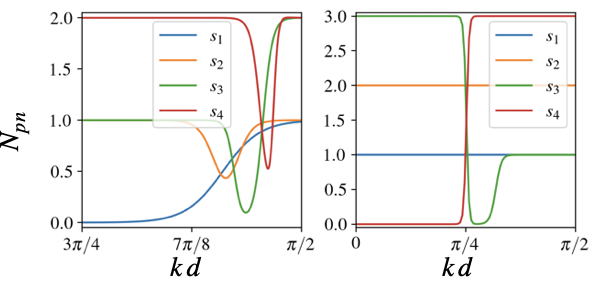}
\caption{The average number of phonons in the the the first four up bands $s_{i}, \ i=1..4$ as a function of polariton momentum is plotted for the systemn with the distance $\varphi=\pi/2$ for two cases: a) $\Gamma_0/\Omega=1/8$ and b) $\Gamma_0/\Omega=5/4$. The second case demonstrates that for narrow resonances inside polariton gap the average number of phonon does not change across the zone and can well describe the corresponding band.}\label{fig:phonon}
\end{figure}

Mixing of bound modes with different numbers of phonons leads to varying the average number of phonons in the mode over the dispersion curve. We calculate the average number of phonons in the band $s_i$, $N_{p}^{(s_i)}$ at momentum $k$ using the Hellman-Feynman theorem {~\cite{Gaurav2024} applying it to the complex energies obtained from the poles of Green function Eq.~\eqref{eq:G}}, yielding
\begin{equation}
    N^{(s_i)}_{pn} = \text{Re}\left[\frac{\partial \omega^{(s_i)}_k}{\partial \Omega}\right].
\end{equation}
{Note that the unitary transformation Eq.~\eqref{eq:U} does not change the observed phonon because it does not contain any explicit dependence on phonon energy.} We plot $N_{p}^{s_i}$ for two sets of parameters in Fig.~\ref{fig:phonon}. It can be seen that in the vicinity of the anticrossing shown in Fig.~\ref{fig:2}(b) the middle mode is a superposition of modes with zero, one and two phonons.

\begin{figure}[t]
\centering
\includegraphics[width=0.48\textwidth]{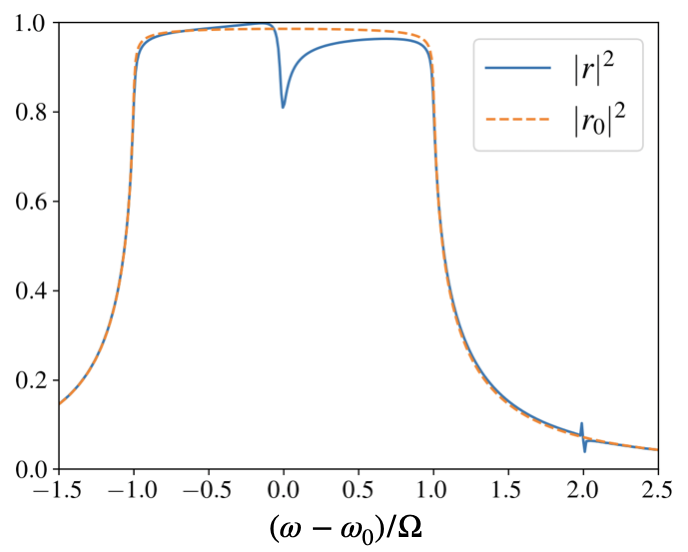}
\caption{Reflection spectrum is plotted as a function of frequency detuning in the case $\Gamma_0/\Omega=1$ and distance $\varphi=\pi/2$ is considered. The dashed line corresponds to the bare polariton case.}\label{fig:refl}
\end{figure}

A straightforward way to probe the impact of phonons in the experiment is to analyze the reflectance spectrum of the system. For this purpose, we study the semi-infinite array of qubits Eq.~\eqref{eq:H_trans2}, which corresponds to $j\geq 0 $.  For more details see the \textbf{Appendix}.  The conception of MAA can be extended to semi-infinite system~\cite{ima}. We thus calculate the phonon dressed  single-particle Green's function which may in turn be used to calculate the reflection coefficient of the system as
\begin{equation}
    r(\omega)=-i\Gamma_0\sum\limits_{l,j}G(i,j,\omega)e^{iqd(j+l)},
    \label{eq:refl}
\end{equation}
where $q$ is the polaritonic momentum. As a result we plot the reflectance spectrum from the semi-infinite system in Fig.~\ref{fig:refl}. As it is predicted, there is a narrow resonance in the polariton band gap and also there are resonances in the vicinity of anticrossing. Away from  these resonances the reflection spectrum repeats polariton spectrum.

It is instructive to analyze the nature of the approximation that led to Eq.~\eqref{eq:H_trans1}. There, we have neglected the fluctuation of the wavenumber $k$ in the dynamical phase induced by the atomic motion $k\hat{z}_j\approx k_0\hat{z}_j$. This is equivalent to the Markov approximation, which neglects the retardation due to the finite time of flight of the photon between individual qubits. The Markovian approximation is valid for not very long arrays, for which $1/\Gamma_0> (Nd)/c$. Thus, while the analysis was performed for infinite structure, as $N\rightarrow\infty$ the retardation effects will have to be taken into account. Making the period of the structure small $\varphi \ll 1$ would allow to suppress the retardation effects.

\section{Conclusions and Outlook}
We report the discovery of a pronounced polaron effect in waveguide quantum electrodynamics, arising from resonant phonon-assisted mixing of polaritons. This interaction introduces new band gaps and in-gap states with weak dispersion, marking a significant advancement in our understanding of optomechanical interactions at the quantum level. The tunability of these effects through inter-atomic spacing provides a novel mechanism for controlling light-matter interactions with cold atoms.

Our findings pave the way for the development of optomechanical quantum memory and nonlinear optics in waveguide QED systems. These results highlight the potential for achieving strong few-photon nonlinearities and point toward new avenues in quantum technologies leveraging hybrid mechanical and photonic systems.



\appendix

\section{Classical noise limit}
\subsection{Model}
\begin{figure}[t]
\centering
\includegraphics[width=0.48\textwidth]{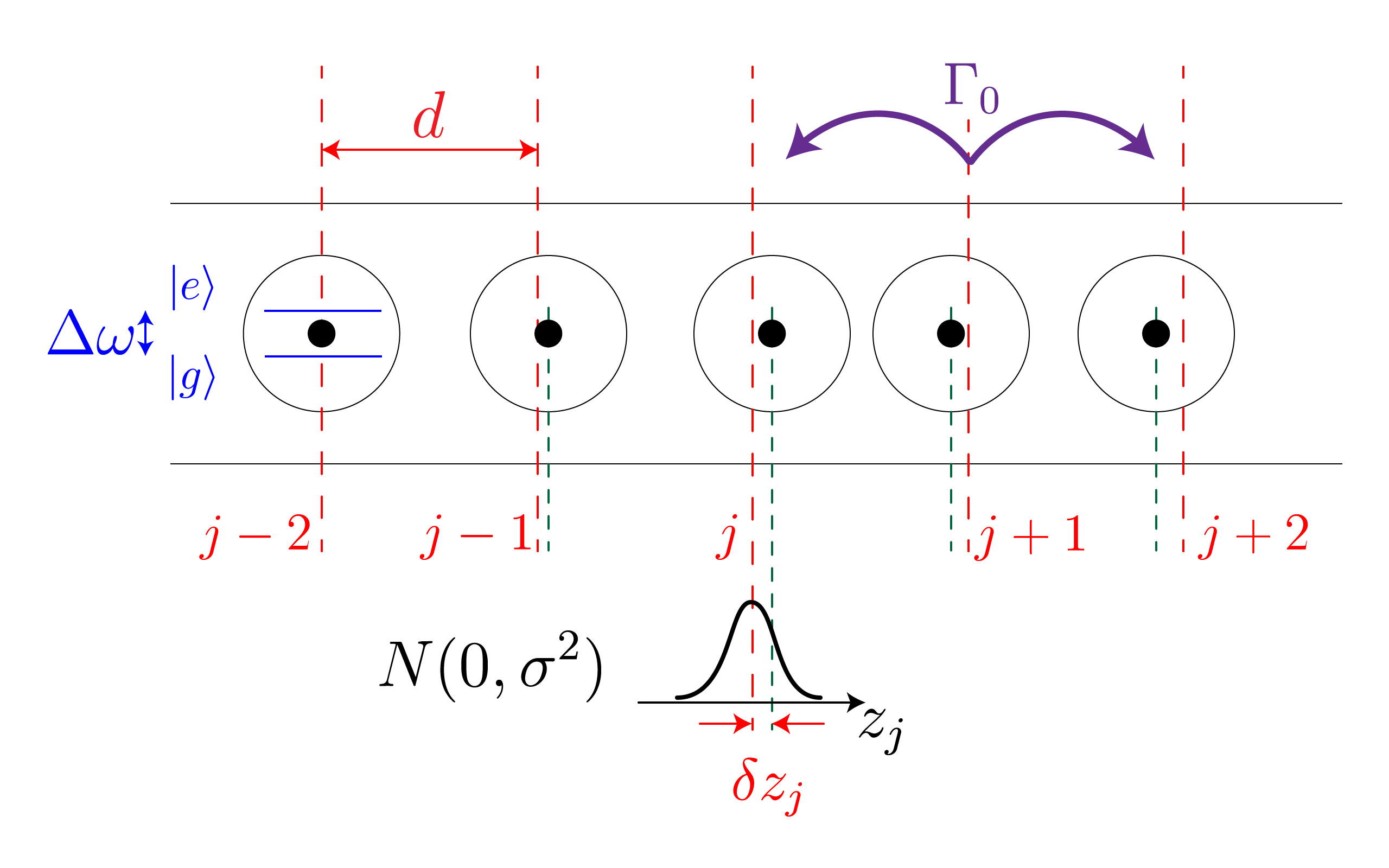}
\caption{Schematic of the semi-classical model.}\label{fig:6}
\end{figure}

In this section, we investigate periodically spaced qubits with mechanical oscillations near their centers in a bidirectional waveguide, modeling qubit displacements using a
classical Gaussian distribution to represent noise Fig.~\ref{fig:6}. This model represents the limit case of quantum mechanical oscillations, when the phonon frequency vanishes, i.e $\Omega\rightarrow0$. So each of the atom position can be represented as 
\begin{equation}
    z_j = jd + \delta z_j, \ \text{where} \ \delta z_j\sim N(0,\sigma^2)
\end{equation}
where $N(0,\sigma^2)$ is the Gaussian distribution with zero expected value and dispersion $\sigma$. The probabilistic nature of the atom positions causes the random strength of exciton hopping from site to site, which makes it possible to  treat them as an effective potential of atom interaction $\hat{H}=\hat{H}_0+\hat{V}$ where
\begin{equation}
\begin{aligned}
    & \hat{H}_0=(\omega_0-i\Gamma_0)\sum\limits_{j=1}^{N}\hat{\sigma}^{\dagger}_j\hat{\sigma}_j; \\&
    \hat{V}=-i\Gamma_0\sum_{l\neq j}^{N}e^{i\varphi|l-j|}e^{i\varphi k_0\text{sgn}_{lj} 
    (z_l-z_j)}\hat{\sigma}^{\dagger}_l\hat{\sigma}_j=\sum_{l\neq j}^{N}t_{lj}\hat{\sigma}^{\dagger}_l\hat{\sigma}_j,
\end{aligned}
\end{equation}
where $t_{lj}$ is a random number with the expected value
So we obtain
\begin{equation}
    \langle t_{lj}  \rangle =-i\Gamma_0 e^{i\varphi|l-j|}e^{-\varphi^2\sigma^2}=-i\Gamma_0 e^{i\varphi|l-j|}\xi=p_{lj}\xi.
\end{equation}
If atoms are totally fixed, i.e. $\sigma\rightarrow0$, then the Gaussian distribution vanishes to delta-type distribution and the old effective model recovers. In the limit of huge noise $(\sigma\rightarrow\infty)$ the expected value of hoping strength vanishes $\langle t_{ij}\rangle\rightarrow0$ and the system transpires in the regime of fully uncoupled atoms. 

\subsection{Transfer matrix formalism}
The main characteristic of such disordered system is the inverse length of localization, mainly because it is possible to measure it in practice. In order to calculate it, the most straightforward method is to use the formalism of transfer matrix. The fundamental theorem in this formalism declares that the transfer matrix of the all system can obtained as consequent multiplication of transfer matrices as light propagates through the system 
\begin{equation}
    \tilde{M}_N=M_NM_{N-1}...M_2M_1,
\end{equation}
where $M_j$ is the matrix that describes how light transforms by passing from $j-1$-th and $j+1$-th atoms through $j$-th. For periodic systems all $M_j$ are the same $M_j=M_0$ for any $j$. Hence, the theorem establishes the fundamental connection between the properties of all system and the properties of each atom, i.e. $\tilde{M}_N=M^{N}_0$. 

In the absence of the noise the system is strictly periodic, and it is naturally to define the elementary cell to introduce a transfer matrix of it. Nevertheless, it is intuitively simple to expand the definition to disorder system by assuming that the coefficients of the transfer matrix can be random. So let the elementary cell has the length $d$ with the center in position of the atom in ordered system, then the transfer matrix for such cell can be written as
\begin{equation}
    M_l=T_{-\delta z_l+d/2}T_{\text{atom}}T_{\delta z_l+d/2},
\end{equation}
where
\begin{equation}
    T_{r} = \begin{pmatrix}
e^{ik_0r} & 0\\
0 & e^{-ik_0r} 
\end{pmatrix},
\end{equation}
is the transfer matrix through the free part of the waveguide with the length r, and 
\begin{equation}
T_{\text{atom}}=\frac{1}{t}\begin{pmatrix}
t^2-r^2 & r\\
-r & 1
\end{pmatrix},
\end{equation}
denotes how the fields from left and right of the atom are linked and expressed via the reflection amplitude $r=-i\Gamma_0/(\omega-\omega_0+i\Gamma_0)$ and transmission amplitude $t=1+r$. Hence, the transfer matrix for an elementary cell is 
\begin{equation}
    M_l=\frac{1}{t}\begin{pmatrix}
(t^2-r^2)e^{i\varphi} & re^{-2ik_0\delta z_l}\\
-re^{2ik_0\delta z_l} & e^{-i\varphi}
\end{pmatrix}.
\end{equation}
As we can see only subdiagonal elements are random numbers. However, in practice only average characteristics can be observed, so it is natural to consider the average transfer matrix
\begin{equation}
    \langle \tilde{M}_{N}\rangle=\langle\prod_{j=1}^{N}M_j\rangle=\prod_{j=1}^{N}\langle M_j\rangle=\langle M_0\rangle^{N},
\end{equation}
where we used the knowledge of statistically independent matrix elements of different cells. So, the expected value for such elementary transfer matrix is 
\begin{equation}
    \langle M_0 \rangle=\frac{1}{t}\begin{pmatrix}
(t^2-r^2)e^{i\varphi} & r\xi^2\\
-r\xi^2 & e^{-i\varphi}
\end{pmatrix},
\end{equation}
which does not depend on particular position of atom.

\subsection{The inverse localization length}

\begin{figure}[b]
\centering
\includegraphics[width=0.48\textwidth]{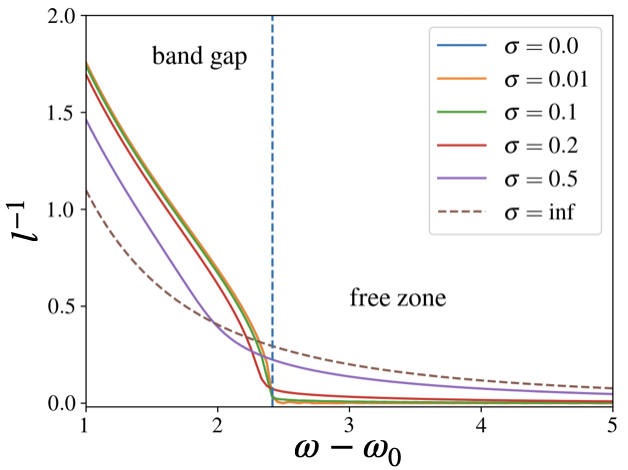}
\caption{The inverse localization length as a function of frequency with fixed noise for case distance $\varphi=\pi/4$. Blue dotted line denotes the edge of the polaritonic band.}\label{fig:7}
\end{figure}

Getting back to the main purpose of using the transfer matrix formalism, we need to establish the connection between the inverse localization length and elements of the global transfer matrix. It can be done with the help of connecting our problem to the condensed matter problem where it can be expressed via the resistance of the chain $R_N$ that has straightforward connection with the transmission coefficient of the chain
\begin{equation}
    l^{-1}=\frac{1}{N} \ln R_N = \frac{1}{N} \ln |T|^{-2}.
\end{equation}

Due to the noise we need to average this equation over disorder
\begin{equation}
    l^{-1}=\frac{1}{N}\langle \ln R_N\rangle \geq \frac{1}{N} \ln \langle R_N\rangle=\frac{1}{N} \ln \langle |T|^{-2} \rangle.
\end{equation}
Due to the inequality in calculating the average transmitted intensity, we will have a consistent evaluation of the inverse localization length. In order to estimate the expected value of the intensity it is important to recall the physical meaning of the matrix elements of the global transfer matrix, i.e . 
\begin{equation}    [\tilde{M}_{N}]_{22}=T^{-1}, \ [\tilde{M}_{N}]_{11}=(T^*)^{-1}.
\end{equation}
Hence, in order to calculate the inverse localization length we need to average
\begin{equation}
    \langle |T|^{-2} \rangle=\langle [\tilde{M}_{N}]_{22}[\tilde{M}_{N}]_{11} \rangle
\end{equation}
Taking into account symmetric group properties ~\cite{pendry1994} this average can be calculated almost analytically. Nevertheless, a simple analytical answer can be calculated for the extremely huge disorder:

\begin{equation}
    l^{-1}=\ln\left[1+\frac{2\Gamma^2_0}{(\omega-\omega_0)^2} \right]
\end{equation}
In Fig.~\ref{fig:7} we plot the inverse localization length as a function of frequency $\omega-\omega_0$ for several different strengths of noise. As it can be seen, finite localization is already observed out of the band gap even in the presence of small disorder.

\section{Reflection spectrum of the semi infinite system}
Inhomogeneous Momentum Average method (IMA) requires the knowledge of the bare polariton Green's function of the system. Finite structures are not translational invariant and so the momentum is no longer a good quantum number. This means that the best way to express the Green's function of the system in real space. According to the Eg.~\eqref{eq:H_trans2} the bare polariton Green's function should be a solution of 
\begin{equation}
    (\omega-\omega_0)G_0(l,j)+i\Gamma_0\sum\limits_{m}G_0(m,j)e^{iqd|l-m|}=\delta_{l,j}
\end{equation}
For the polariton in semi-infinite system we have
\begin{equation}
    G_0(l,j,\omega) = \frac{\delta_{l,j}}{\Delta\omega}+Ae^{iqd|l-j|}+r_{\infty}Ae^{iqd(l+j)},
    \label{eq:demiinf}
\end{equation}
where $\Delta\omega=\omega-\omega_0+i\eta$, $A=-i\frac{\Gamma_0}{\Delta\omega^2}\frac{\sin(\varphi)}{\sin(qd)}$ is the amplitude factor, $r_{\infty}=-(1-e^{-i(\varphi-qd)})/(1-e^{-i(\varphi+qd)})$ and $q$ - is the polaritonic momentum that can be find as
\begin{equation}
    \cos(qd)=\cos(\varphi)+\frac{\Gamma_0}{\Delta\omega}\sin(\varphi).
\end{equation}
The first two parts in Eq.~\eqref{eq:demiinf} is the Green's function part of the infinite part, namely $\int dk G(\omega,k)e^{-ik|l-k|}=\frac{\delta_{l,j}}{\Delta\omega}+Ae^{iqd|l-j|}$. The last part is the contribution of the border, namely reflection from the border back inside the structure. 

To calculate the reflection amplitude we can simplify the expression Eq.~\eqref{eq:refl}
as
\begin{equation}
\begin{aligned}
&s^{+}_0(l)=\sum\limits_{j}G_0(l,j)e^{iqdj} \\&=\frac{-i}{\Gamma_0}\left[\delta_{j,0}-(\omega-\omega_0)G_0(j,0)\right],
\end{aligned}
 \label{eq:s_plus}
\end{equation}
where we use the reciprocity of Green's function. Finally we get
\begin{equation}
\begin{aligned}
    r_0(\omega) &= \frac{1}{i\Gamma_0}\left[(\omega-\omega_0-i\Gamma_0)-(\omega-\omega_0)^2G_0(0,0)\right] \\ &=-i\frac{(\omega-\omega_0)}{\Gamma_0}(1-e^{i(qd-\varphi)})-1.
    \end{aligned}
\end{equation}

Now it is possible to calculate directly the Green's function for phonon-polariton model as
\begin{equation}
    \begin{aligned}
        G(l,j,\omega) &= G_0(l,j,\omega')\\&+\sum\limits_{j_1}G(l,j_1,\omega)v_{j_1}(\omega)G_0(j_1,j,\omega'),
    \end{aligned}
\end{equation}
where $\omega'=\omega - \Sigma_{\text{MAA}}(\omega)$, which in turn can be calculated from Eq.~\eqref{eq:Sigma}, $v_j(\omega)=\Sigma_{\text{IMA}}(j, \omega)-\Sigma_{\text{MAA}}(\omega)$ and $\Sigma_{\text{IMA}}(j,\omega)$ has the same structure as in Eq.~\eqref{eq:Sigma} with replacement averaged infinite Green's function to $G_0(j,j,\omega)$.

One may notice that summing the row straightforwardly in Eq.~\eqref{eq:refl} is a complicated numerical problem. Instaed we can use the formalism of $s^{+}$ for the Green's function as in Eq.~\eqref{eq:s_plus}. So it is possible to rewrite Eq.~\eqref{eq:refl} in terms of $s^{+}$ as
\begin{equation}    r(\omega)=r_0(\omega')+\sum\limits_{j}s^{+}(j,\omega)v_j(\omega)s^{+}_0(j,\omega'),
\end{equation}
where $s^{+}_{j}$ can be found as
\begin{equation}
    s^{+}(j,\omega)=s^{+}_0(j,\omega')+\sum\limits_{j_1}s^{+}(j_1,\omega)v_{j_1}(\omega)G_0(j_1,j,\omega'),
\end{equation}
which can solved effectively, because the amplitude of $|v_j|$ decreases rapidly on all frequency range except the vicinity of band edges, where it decreases slower. We know spectral weights in points that are far from the border should be the same as in MAA, namely 
\begin{equation}
\begin{aligned}
A(j,j,\omega)&\rightarrow-\frac{1}{\pi}\text{Im}\frac{1}{N}\sum\limits_{k}\frac{1}{\omega-\epsilon(k)-\Sigma_{\text{MAA}}(\omega)} \\ & -\frac{1}{\pi}\text{Im}\left[\bar{g}_0(\omega')\right].
\end{aligned}
\end{equation}
This limit can be used as a criteria of the sufficient number of iterations in the sum that can reach the appropriate accuracy. 

\bibliography{Polaron.bib}

\end{document}